\documentclass[reprint, amsmath,amssymb,groupedaddress]{revtex4-1}

\usepackage{graphicx}
\usepackage{dcolumn} 
\usepackage{bm}%
\usepackage{multirow}
\begin{document}


\title{Chain structure of head-on collisions in boundary driven granular gases}


\author{Yanpei Chen}
\email[]{ypchen@ipe.ac.cn}
\affiliation{State Key Laboratory of Multiphase Complex Systems, Institute of Process Engineering, Chinese Academy of Sciences, Beijing 100190, China}
\author{Wei Wang}
\email[]{wangwei@ipe.ac.cn}
\affiliation{State Key Laboratory of Multiphase Complex Systems, Institute of Process Engineering, Chinese Academy of Sciences, Beijing 100190, China}
\affiliation{University of Chinese Academy of Sciences, Beijing 100049, P.R. China}
\author{Meiying Hou}
\affiliation{
 Key Laboratory of Soft Matter Physics, Beijing National Laboratory for Condense
Matter Physics,Institute of Physics, Chinese Academy of Sciences, Beijing 100190,
China}
\affiliation{
College of Physics, University of Chinese Academic of Sciences, Beijing 100049,
China
}%

\date{\today}

\begin{abstract}
We report a peculiar dynamic phenomenon in granular gases, chain structures of head-on collisions caused by the boundary heated mechanism form a network in an Airbus micro-gravity experiment and horizontal vibrated one in the laboratory, which differ markedly from the grazing-collision-dominant in randomly driven granular fluid. This new order property is an orientation correlation between the relative position and the relative velocity of any particle pair, which weakens the collision frequency and leads a long range boundary effect. By the histogram of the relative position and the relative velocity, we find this position-velocity  correlation is not only at limits of very small relative velocities but also large ones, which means the breakdown of molecular chaos assumption is not limited to a small portion of the phase space\cite{pagonabarraga2001randomly}. Through a simple  anisotropic angular distribution model of the relative position and the relative velocity, we could modify classical uniform angular integration results of mean field values taking the effect of the observed collision chain structure explicitly into account.

\end{abstract}


\maketitle

Rapid granular flows, or granular gases, needing continuous energy input to balance dissipation, hitherto show a large amount of interesting out-of-equilibrium phenomena. In the real world, energy is injected generally by means of the boundary vibration, shear or others. In the case of a granular gas confined between two vibrating walls without gravity, the system form a spatially gradient fields with the distance of the energizing boundary. What's more, there are gradual velocity distributions\cite{brey2000boundary,herbst2004local,ISI:000327241500003}( from two-peak near the boundary layer to one peak in the center layer), and abnormal large local mean free paths\cite{mei2016the} near the boundary. These represent a long range boundary effect\cite{yanpei2011long,Menon2008Heating}. It is apparent from gradual fields that the heating boundary brings a new characteristic length scale\cite{vollmayrlee2011hydrodynamic}, which is difficulty to predict by the classical kinetic theory, hydrodynamic equations\cite{brey2000boundary} and hydrodynamic fluctuation theory \cite{prados2011large,Otsuki2008Spatial}. So the boundary effect are usually seldom touched\cite{puglisi2005fluctuations,vollmayrlee2011hydrodynamic}, however, boundary effect can not be ignored\cite{PhysRevLett.100.158001,PhysRevLett.99.028001} for granular gases.

 Identifying and quantifying the boundary scale effect is the accepted prerequisite before avoiding this effect. Then if the influence of the side walls is able to be considered, and the classical kinetic theory and hydrodynamic equations could be modified further. To counter the problems above, deeply analyzing the boundary length scale, or correlation (the breakdown of \emph{molecular chaos}\cite{goldhirsch2003rapid,pagonabarraga2001randomly}) is critical, which also helpful in interpreting about the attractors in the phase space\cite{grossman1997towards}. Although randomly driven granular gases demonstrate correlation in both experiment\cite{PhysRevLett.89.084301} and numerical simulations\cite{pagonabarraga2001randomly}, moreover, quantitative predictions of short-range correlation\cite{pagonabarraga2001randomly} and long-range correlation\cite{PhysRevE.59.4326} are given by the mode coupling theory and in the frame work of isotropic hydrodynamics, the orientational correlation\cite{brilliantov2007translations,PhysRevE.89.062201,gayen2008orientational} has been found anisotropic in an uniformly heated system, and is said to be responsible for the emergence of non-Gaussian high-velocity tails. In contrast to the well-know physical property of clustering, we still lack a clear picture of microstructure of a long boundary effect. Two-peaks (non-Gaussian) velocity distributions\cite{brey2000boundary,herbst2004local,ISI:000327241500003,yanpei2012breakdown} imply correlated velocities\cite{prevost2002forcing,baxter2007experimental}.  In addition, we notice that the boundary shape maybe affect the field\cite{PhysRevLett.118.198003}, therefore we focused on the orientational correlation between the relative position and the relative velocity here, which maybe one of the key source to understand above the long boundary effect. Furthermore, it is of interest to ascertain the impact of such anisotropic orientational correlation on the mean field values, which could be helpful for the non-equilibrium dynamic description.

  In this letter, we report a peculiar dynamic phenomenon in granular gases---chain structures of head-on collisions form a network, caused by the boundary heated mechanism. This is an orientation correlation between the relative position and the relative velocity by analyzing micro-gravity experimental data and horizontal vibration conditions. In detail, head-on collisions prevail for the boundary heating granular gases, which is entirely different from grazing collision in randomly driven granular fluids\cite{goldhirsch1993a,Tan1997Intercluster}. This means most of angles between $\bm{c}_{ij}$ and $\hat{\bm{k}}$ are around zero or $\pi $\cite{soto2001statistical} in our system, where $\hat{\bm{k}}$ is the unit vector directed from the center of particle $i$ to that of $j$, and their relative velocities, $\bm{c}_{ij}$ ($\bm{c}_{ij}=\bm{c}_{i}-\bm{c}_{j}$). If connecting particles involved in head-on collision, we could find a chain microstructure between two heating boundaries, like the force chain between two shear boundaries in shear granular solids. This is what call the collision chain, which is a new long range structure in granular gases. The existence of the collision chain could make the collision frequency lower than the homogeneous cases, which goes against with  previous results that suggested by structures, for instance, clusters could make it higher. Using anisotropic angular distribution between the relative position and the relative velocity, we are able to describe this correlation in granular gases through a new parameter, angular factor, $\beta_{m}$. Compared with the Enskog's factor,  $\beta_{m}$ could account the orientation correlation between the relative position and the relative velocity.

-\emph{Experiment} We investigated 2D vibro-fluidized experiments combing two results under two environments: one is micro-gravity Airbus of Novespace (2006 Campaign) (denoted by $\mathbf{A}$), the other is in laboratory but by a horizontal vibration(denoted by $\mathbf{B}$). There is no effect of gravity in both of $\mathbf{A}$ and $\mathbf{B}$. Experimental parameters are listed in Table \ref{tab:table1}.  The detailed experimental settings of micro-gravity of $\mathbf{A}$ one could be found in the previous work\cite{chen:tel-01141249,yanpei2012breakdown}. There are 47 bronze beads accompanied by various vibration strength in  $\mathbf{A}$, the area fraction is 0.54£¬while in $\mathbf{B}$ the particle number ranges from 16 to 272 under the same vibration. In micro-gravity experiment, it is difficult for us to repeat the experiments in various particle numbers, so $\mathbf{B}$ can be viewed as compensatory case. The movements of particles are recorded by using a high-speed camera (499/500 frames per second in both $\mathbf{A}$ and $\mathbf{B}$).  We ignore the sliding fraction in $\mathbf{B}$.

\begin{table}
\caption{\label{tab:table1}Summary of experimental parameters. $D$ is the diameter of particle, $L$, $W$ and  $H$ are length,  width and  height of the cell.  $R$ is the spatial resolution of high-speed camera, given in pixels. $f$ is frequency, $\gamma$ is the vibrational acceleration.}
\begin{ruledtabular}
\begin{tabular}{l|c|ccc|cc|cc}
\multirow{2}{*}{\textrm{Experiments}}&\multirow{2}{*}{D($mm$)}&\multicolumn{3}{c|}{cell($mm$)}&\multicolumn{2}{c|}{R(pixels)}&\multicolumn{2}{c}{vibration}\\
&&$L$&$W$&$H$&$L$&$W$&$f$(Hz)&$\gamma$($m/s^{2}$)\\ \hline
Micro-gravity& 1.21 &  10 & 10&  1.4  &288 &288&-&-\\
Horizontal vibration   & 3 &  70&  50&  10&  880&650&60&124\\
\end{tabular}
\end{ruledtabular}
\end{table}

\begin{figure}
 \includegraphics[width=3in]{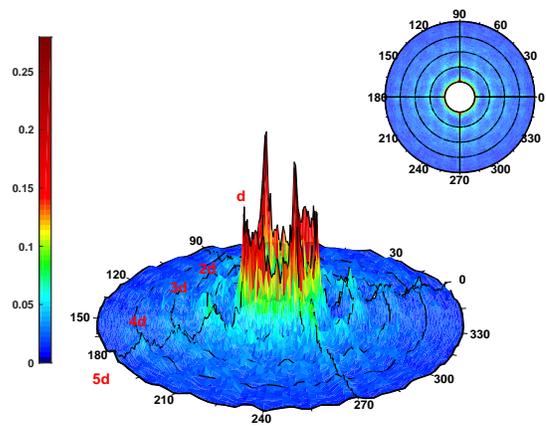}
 \caption{\label{angluarpaircorrelation}The radial-angular correlation function $g(r, \theta)$ in micro-gravity, for which parameters are listed in Table \ref{tab:table1} ($\mathbf{A}$), and the frequency of vibration is 49 $Hz$, and acceleration is 21.6 $m/s^{2}$.}
 \end{figure}

In Fig. \ref{angluarpaircorrelation}, we firstly present the radial-angular correlation distribution of particles in micro-gravity $\mathbf{A}$, $g(r, \theta)$, namely the density-density correlation function. In spite of resembling a liquid-like structure(Radial distribution function $g(r)$ is not shown here), $g(r, \theta)$ shows two spikes along the vibrating direction, indicating anisotropy. This anisotropy is similar to the collision layer of Leidenfrost state\cite{ansari2016pattern-transition} in gravity environment. However, to identify the orientational correlation, we center our attention on study of three relevant parameters: (1) the direction probability of the relative position $\hat{\bm{k}}$ between any two particles, $P(\hat{\bm{k}})$, (2) the probability of their relative velocity  $\bm{c}_{ij}$, (3) the direction probability of $\bm{c}_{ij} \cdot \hat{\bm{k}}$, as discussed in the following parts, respectively. All of these parameters are isotropic or uniform in classical kinetic theory.

\begin{figure}
\includegraphics[width=3in]{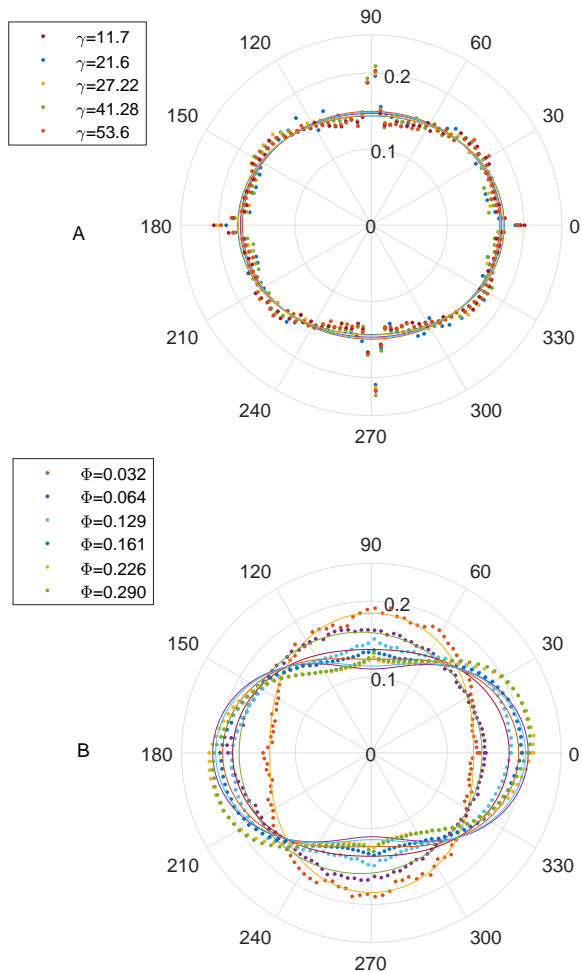}
\caption{\label{fig:angk0} The dependence of $P(\bm{k})$ of $\mathbf{A}$: on vibration ($\gamma
(m/s^{2})=\ 11.7, 21.6,\ 27.22,\ 41.28,\ 53.6$, $f (Hz)=49,\ 49,\ 97,\ 97,\ 97$), and $\mathbf{B}$ on area fraction ($\Phi=0.032,\ 0.064,\ 0.129,\ 0.161,\ 0.226,\ 0.290$) under the polar coordinate. The tilted degree $D=0.5^{\circ}$ in $\mathbf{B}$. The rest parameters could be found in Table \ref{tab:table1}.}
\end{figure}

The orientational distributions $P(\hat{\bm{k}})$ of $\mathbf{A}$ and $\mathbf{B}$ are shown in  Fig. \ref{fig:angk0}, where $\hat{\bm{k}}$ is the unit vector of two arbitrary particle centers. $P(\hat{\bm{k}})$ clearly displays heterogeneous, which is totally different from uniform molecular gases, $P(\hat{\bm{k}})=1/\pi$. $P(\hat{\bm{k}})$  is flattened into an oblong shape along y axis (the vibration direction), and the anisotropy increases proportionally with the area fraction. Nevertheless, $P(\hat{\bm{k}})$ is not sensitive enough to the vibration acceleration. It suggests that the orientation of the relative position is only affected by the number density of particles, not by the boundary vibration strength, which is very interesting. $P(\hat{\bm{k}})$ can be approximated by a truncated Fourier expansion as \cite{alonsomarroquin2005role}:
\begin{equation}
  P(\alpha)= 1/2\pi \{1+a_{1}\cos( \alpha) + a_{2}\cos(2 \alpha)\}
     \label{Psi}
\end{equation}
where $a_{1}$ and $a_{2}$ could be viewed as anisotropy parameters, corresponding to $\cos  \alpha$ and $\cos 2 \alpha$, respectively. The result of this coupling model fits the measured value very well. The only a few of deviations from the curves are points close to  $\alpha=0$, $\pi/2$, $\pi$ and $3\pi/2$ in $\mathbf{A}$. The reason is still unclear but it presents an orientational order.

\begin{figure}
 \includegraphics[width=3in]{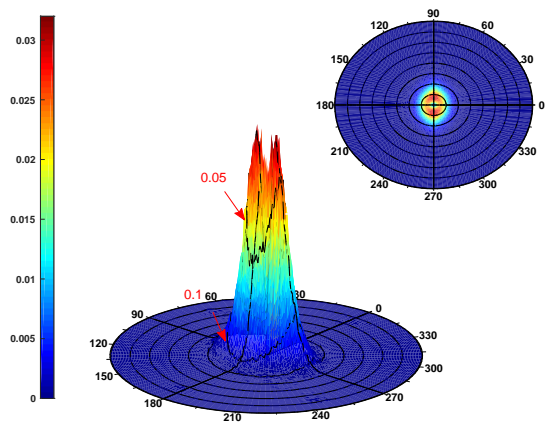}
 \caption{\label{c12distribution} The distribution function of micro-gravity $\mathbf{A}$. Experimental parameters are listed in the Table \ref{tab:table1}, and the frequency of vibration is 49 $Hz$, and acceleration is 21.6 $m/s^{2}$.}
 \end{figure}

Furthermore, the orientation distribution $P(\bm{c}_{ij})$ of $\mathbf{A}$ in plotted in Fig. \ref{c12distribution}, where $\bm{c}_{ij}$ is the relative velocity between two arbitrary particles. Results of $\mathbf{B}$ are similar to those of  Fig. \ref{c12distribution} and not displayed here. For a molecular gas, $P(\bm{c}_{ij})$ is supposed to be isotropic and have a distribution $\sqrt{(2/\pi)} c_{ij}^{2}e^{-1/2c_{ij}^{2}}$\cite{pagonabarraga2001randomly}. However, $P(\bm{c}_{ij})$ in our cases is anisotropic and oval shaped, with the maximum of relative velocity $\bm{c}_{ij}$ along $y$ axis, the vibration direction. This is due to fact: particles gain the maximum speed along $y$ axis ($\pi/2$ or $3\pi/2$) after collision with the boundary, then in the center of the cell, the velocity of particles became more isotropic and smaller and the number density is maximum, so the maximum value of $P(\bm{c}_{ij})$  is correspond to the relative velocity between particles at the center layer and boundary layers. Their directions are around $\pi/2$ or $3\pi/2$ along $y$ axis.

\begin{figure*}
\includegraphics[width=5in]{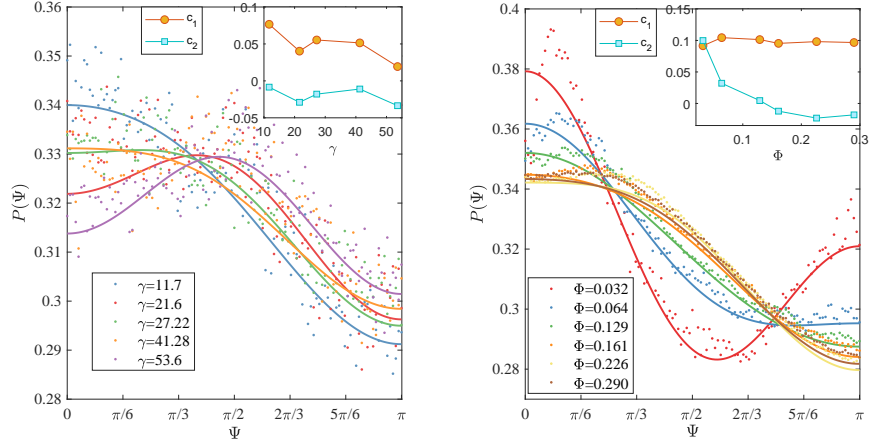}
\caption{\label{ang_ck_all} The orientation distribution, $P(\Psi)$ of $\mathbf{A}$ (Left) and $\mathbf{B}$ (Right),  $\Psi$ is angle between the relative velocity and the relative position. The dashed line is obtained from experiment results, while the solid line is predicted value by  Eq. \ref{Psi2}. Inset: The fitting coefficients corresponding to Eq. \ref{Psi2} versus the acceleration and volume fraction.}
\end{figure*}

Now we turn to the orientation distribution $P(\Psi)$ , where $\Psi$ is the angle between the relative velocity and the relative position, $\cos \Psi\equiv \bf{c}_{ij} \cdot \hat{\bf{k}}/|c_{ij}|$. We know that if $\cos \Psi$ is positive then particles move away from each other(post-collision states), and if $\cos \Psi$ is negative, particles move in close(pre-collision states). Moreover, $\Psi$ is related to the proportion between with the correlations of longitudinal and transverse velocities correlations $c_{\parallel}$  and $c_{\perp}$\cite{PhysRevLett.89.084301}. For a fluidized granular fluid, Soto and Mareschal \cite{soto2001statistical} derived a relation between the post- and pre- collision radial distribution functions at contact as a function of $\Psi$, $[\cos(\Psi)^{2}+\alpha^{2}\sin(\Psi)^{2}]^{-1}$, $\alpha$ is the restitution coefficient, but their model still takes isotropic $\Psi$ of the pre-collision as an ansatz. Here, we plot $P(\Psi)$ of $\mathbf{A}$ and $\mathbf{B}$ in Fig. \ref{ang_ck_all}. Clearly $P(\Psi)$ is not uniform, furthermore, increasing $\gamma$ or $\Phi$ lead to a flatter and more normal curve until there seems to be two plateaus. The proportion of post-collision state (0 $\leq$ $\Psi$ $ \leq $$\pi/2$) is larger than that of pre-collision states ($\pi/2$ $\leq \Psi$$  \leq$$ \pi$). This point is similar with the Enskog's factor $\chi$ derived by Soto and Mareschal\cite{soto2001statistical} ($[\cos(\Psi)^{2}+\alpha^{2}\sin(\Psi)^{2}]^{-1}>1$ when $\alpha<1$). However, our pre-collision states is not a constant which is different with the previous literature\cite{soto2001statistical}. It is need to note that $P(\Psi)$ here account all particles pairs in our experiment, rather than $\Psi$ near one particle diameter at contact. Hence, $\Psi$ here is related to the number density and the velocity, not a generalized pair correlation function at contact. We could also fit it using a Fourier expansion,
\begin{equation}
    P(\Psi)= 1/\pi \{1+c_{1}\cos( \Psi) + c_{2}\cos(2 \Psi)\}
     \label{Psi2}
\end{equation}
 The fitting results are plotted in the inset of Fig. \ref{ang_ck_all}. Similarly, $c_{1}$ and $c_{2}$ could also be seen as the anisotropy parameters. The dependence on the acceleration and volume fraction of $c_{1}$ and $c_{2}$  are also shown in the inset. Non-zero value of  $c_{1}$ and $c_{2}$ demonstrate clearly $P(\Psi)$ is anisotropic.
\begin{figure}[htbp]
 \includegraphics[width=3in]{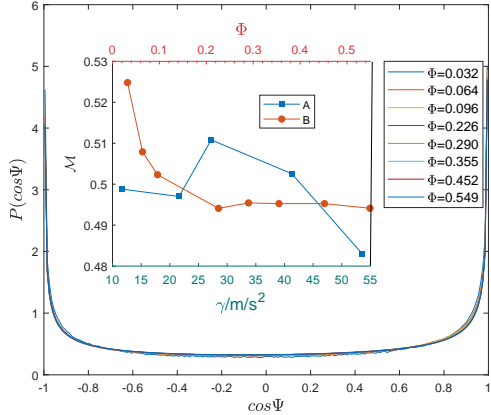}
 \caption{\label{Pcosnuma} The probability distribution of $\cos \Psi'$ for $\mathbf{B}$ with vibrating parameters shown in Table \ref{tab:table1}. The inset shows $\cal{M}$ measured as a function of area fraction $\Phi$ or $\gamma$ for $\mathbf{A}$ and $\mathbf{B}$.}
 \end{figure}
To obtain more detailed information of the orientational correlation between the relative velocity and the unit vector of the line of center,  we can quantify them by the mean square of the cosine of the angle, $\Psi'$,
  \begin{equation}
{\cal M }=\frac{1}{N}\sum _{i=1}^{N}\frac{((\bm{c}_{ij}-\langle \bm{c}_{ij}(t)\rangle)\cdot (\hat{\bm{k}}-\langle\hat{\bm{k}}(t)\rangle))^{2}}{c_{ij}^{2} k^{2}} =\frac{1}{N} \sum_{i=1} ^{N} \cos ^{2}\Psi'
  \label{correlation}
 \end{equation}
where $\langle \rangle$ is the average at time $t$ by counting up all particles in each frame, N is total number of particles of all frames. If there is no correlation between the relative position and the relative velocity, as in 2D gases, one obtains $\cal M $$=1/2$. The evolution of $\cal{M}$ with the volume fraction $\Phi$ for $\mathbf{A}$ and $\mathbf{B}$ is plotted at right of Fig. \ref{Pcosnuma}. $\cal{M}$ in both of $\mathbf{A}$ and $\mathbf{B}$ deviates from $1/2$, which demonstrates that the relative position and the relative velocity correlation exists in vibro-fluidized granular gases. $\cal{M}$  decreases monotonously with increasing of volume fraction $\Phi$, and remains relatively stable till $\Phi=0.2$ for $\mathbf{B}$.

 To examine more closely, we also plot the probability distribution of $\cos \Psi'$ in $\mathbf{B}$ in left of Fig. \ref{Pcosnuma}. There are two peaks clearly located at $-1$ and $1$ for $P(\cos \Psi')$, respectively, in our system. It is obvious the probability of head-on collisions is higher than that of the oblique one, which is consistent with above results and our previous event-driven molecular dynamic simulation results \cite{mei2016the}. It need to mention that $P(\cos \Psi')$ between translational and rotational velocities \cite{gayen2008orientational} in uniform shear flow has a peak at $0$, which is reasonable because that the uniform case is dominated by grazing-collision.





\begin{figure}
 \includegraphics[width=3in]{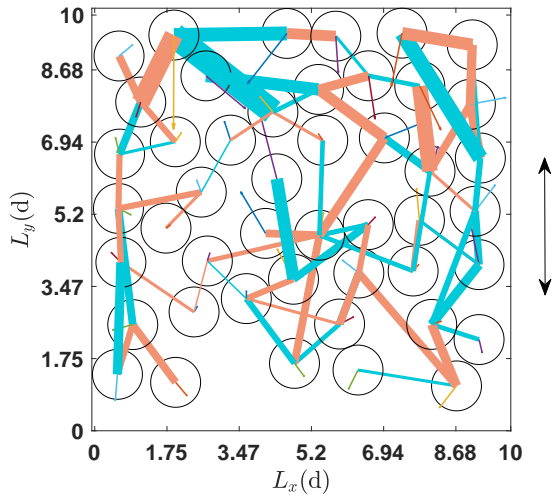}
 \caption{\label{collisionchain} Collision chains in the micro-gravity $\mathbf{A}$. The experimental parameters are the same as in Fig. \ref{angluarpaircorrelation}. The light-coral-lines-connected particles satisfy $\cos \Psi' \subset ( [-1 \quad -0.9] )$, The cyan lines for $\cos \Psi' \subset ( [0.9 \quad 1] )$ and their width are proportional to their relative velocities. The arrow line is the particle velocity. The vibration direction is shown here via the double-headed black arrows. $d$ is the particle diameter. }
 \end{figure}

 Inspired by Fig. \ref{collisionchain}, we draw lines between pairs of particles which satisfy that $\cos \Psi' \subset ([-1 \quad -0.9] \cup [0.9 \quad 1] )$, i. e., $\Psi' \subset ([0^{\circ} \quad 25^{\circ} ] \cup [155^{\circ}  \quad 180^{\circ} ] $, and the distance of a pair of particles is the nearest for the reference particle with all the other particles. In addition, the width of the line is drawn proportional to their relative velocity. As is vividly depicted in the drawing, most particles have head-on collision relationships with their adjacent particles. There seems to be chain structures like force chains in granular solids, connecting one particle to another between two driven boundary, though the whole system seems homogenous. Let us define a collision chain consists of a set of particles with in a boundary driven rapid granular material that are held together and  trapped by a network of head-on collision. From Fig. \ref{collisionchain}, we could find these collision structure likely govern the system's giant number particles.

 We elucidate this phenomenon by the correlation\cite{soto2001statistical} between the relative position and the relative velocity affected by the boundary heating. After colliding with the heating boundary mechanism, the particle gain the velocity pointing in the direction of the axis $y$. Later on, because of correlation, the post-collision relative velocities of a particle collision pair become smaller, and more parallel than in the elastic case. That means the particle movements are more close to the movement with the velocity pointing in the direction of the axis $y$. Eventually, it is convenient to form a chain from the boundary layer to the center after serval inelastic collisions. this demonstrates even the dilute granular gases without cluster could emerge dynamic structures which making it far from equilibrium.

In order to quantify and introduce anisotropic orientation impacts on granular behavior into kinetic theory, a simple but very practical method is to define a dimensionless angular integral, named angular factor,
\begin{equation}\label{angularinter}
\beta_{m}=\frac{\mathcal{J}_{m}}{\mathcal{J}_{m}^{E}}=\frac{\pi\int d \hat{\bm{k} }\Theta (-\hat{ \bm{k}} \cdot \hat{ \bm{g}})(\hat{ \bm{k}} \cdot \hat{ \bm{g}})^{m} P(\Psi) }{\int d \hat{\bm{k} }\Theta (-\hat{ \bm{k}} \cdot \hat{ \bm{g}})(\hat{ \bm{k}} \cdot \hat{ \bm{g}})^{m} }
\end{equation}
where $\hat{\bm{g}}\equiv \bm{c}_{ij}/c_{ij}$ is the unit vector directed along $\bm{c}_{ij}$, $P(\Psi)$ is the probability density distribution of $\Psi$, $\Theta$ is  the Heaviside step-function.  The denominator is the integral result corresponding to the evenly orientational distribution\cite{brilliantov2010kinetic},
\begin{equation}\label{je}
  \mathcal{J}_{m}^{E}=\pi^{\frac{1}{2}}\frac{\Gamma (\frac{m+1}{2})}{\Gamma (\frac{m+2}{2})}
\end{equation}

\begin{figure}[htbp]
 \includegraphics[width=3in]{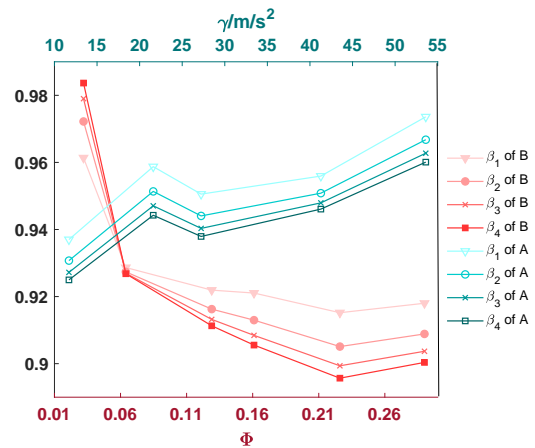}
 \caption{\label{chao1} The dimensionless angular integral $\beta_{m}$ as a function of the area fraction and acceleration of $\mathbf{A}$ and $\mathbf{B}$.  }
 \end{figure}

 It is known that collision frequency, pressure and the energy dissipation contain the factor $|\bm{c}_{ij} \cdot \bf{k}|^{m}f^{(2)}(\bf{c}_{i},\bf{c}_{j},\bf{k})$, corresponding to the case of $m=1,\,2,\,3$, where $f^{(2)}(\bf{c}_{i},\bf{c}_{j},\bf{k})$ is the dynamic or constrained pair distribution function velocities. So $|\cos \Psi|^{m}$ is used to investigate the breakdown of the molecular chaos assumption. For randomly driven dissipated granular fluids\cite{pagonabarraga2001randomly}, the simulation results show that $\beta_{m}>1$, and larger $m$, closer to unity $\beta_{m}$ is. In our case, we can introduce the anisotropy of $\Psi$ (Eq. \ref{Psi2}) into $\beta_{m}$,  by applying our fitting parameters of $P(\Psi)$ of Eq. (\ref{Psi2}) to Eq. (\ref{angularinter}). Fig. \ref{chao1} illustrates $\beta_{m}$ of $\mathbf{A}$ and $\mathbf{B}$. We could find that, $\beta_{m}$ begins to deviate from unity, but, $\beta_{m}<1$ in our case. Fig. \ref{chao1} demonstrates that, in boundary vibro-fluidized granular gases, $|\cos(\Psi)|^{m}$ in the collision frequency is smaller than ones in homogeneous state, so is the energy dissipation. We know that the Enskog factor enhance the collision frequency at higher density, however, this new structure appeared, the collision chain make $\beta_{1}<1$, that means it weakens the collision frequency. This is reason why the local mean free path near the boundary is longer than the homogeneous theory prediction, which cause a long range boundary effect.

 Furthermore, our results show that larger $m$, more deviation to unity  $\beta_{m}$ is. Deviation increases with the area fraction, and decreases with the acceleration of vibration. This occurs because that larger vibration or diluter density make the system more homogeneous, then $\beta_{m}$  is closer to 1.  It is reasonable that most collisions in our cases are `` head-on'', while most collisions in randomly driven are grazing ones.

The variation of $\beta_{m}$ implies we could improve the kinetic theory by introducing different anisotropy $P(\Psi)$ under various boundary shapes. Compared with the Enskog's factor\cite{soto2001statistical} method, $\beta_{m}$ can not only account the spatial correlation but also the position-velocity correlation. Especially in our system, two-peak velocity distribution deviated distinctly from the Gaussian distribution, make that the position-velocity correlation can not be ignored. Merely using of anisotropic orientation distribution we could describe this structure.

\begin{figure}
 \includegraphics[width=3in]{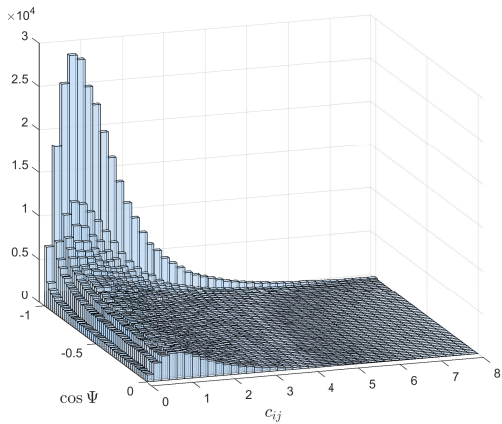}
 \caption{\label{molecular} The histogram of the $c_{ij}$ and $\cos \Psi$ of  $\mathbf{B}$ with $\Phi=0.226$. }
 \end{figure}

 Figure \ref{molecular} shows the  histogram of $c_{ij}$ and $\cos \Psi$ which is limited in the range of $\cos \Psi \subseteq [-1 \quad  0]$ and the relative distance $|\textbf{r}_{ij}|<2d$. These data could be viewed as belonging to the pre-collision state. For the randomly driven granular fluids, one argument\cite{pagonabarraga2001randomly} is that the molecular chaos assumption only breaks down in a very small portion of the phase space, where $|c_{ij}|$ is small. In our system, most of data are located at $\cos \Psi \rightarrow -1$, not $\cos \Psi \rightarrow 0$. The phase space in the boundary heating granular gases is totally different with randomly driven ones. That is in our cases, it is hard to support that the molecular chaos only breaks down in a small relative velocity space. It is clearly that the boundary heating mechanism leads the phase space change.

In conclusion, we found a new phenomenon in boundary heating granular gases, chains structure of head-on collisions form a network between two driven boundaries as force chains in shear granular solids. Unlike cluster structure, chains structure weakens the collision frequency and  is the most likely explanation of a long range boundary effect. By introducing the anisotropy of the relative position and the relative velocities orientational distribution, we can introduce this orientation correlation to mean field values. The quantitative results of the boundary scale have not yet been obtained here and need further investigation, however, we give a convincing explanation why the boundary heating mechanism can not be ignored in inelastic gases\cite{PhysRevLett.100.158001,PhysRevLett.99.028001}. So it have significance to perfecting the kinetic theory and hydrodynamic theory of dilute granular flows. Applications to industry boundary design of the process and transport of the granular materials may also be considered.

The authors acknowledge funds from the National Natural Science Foundation of China under Grant Nos. 91334204, 11702291, 11474326, U1738120 and 21625605.
\bibliography{apssamp}

\end{document}